\documentclass[sigconf]{acmart}
\usepackage[utf8]{inputenc}
\pdfoutput=1 

\usepackage{url}
\usepackage{latexsym}
\usepackage{amsmath, amsfonts}
\usepackage{algorithm, algorithmic}
\usepackage{graphicx}
\usepackage[binary-units]{siunitx}
\usepackage[textsize=tiny]{todonotes}

\usepackage{pgfplots}
\usepgfplotslibrary{groupplots,dateplot}
\usetikzlibrary{patterns,shapes.arrows}
\pgfplotsset{compat=newest}
\usepackage{hyperref}
\usepackage{booktabs}
\usepackage{subfig}
\usepackage{tabularx}
\usepackage{csquotes}

\newcommand{\tildeSymbol}{{\raise.17ex\hbox{$\scriptstyle\sim$}}}

\usepackage{xspace}
\newcommand*{\eg}{e.g.\@\xspace}
\newcommand*{\ie}{i.e.\@\xspace}

\begin{document}

\sloppy

\copyrightyear{2022} 
\acmYear{2022} 
\setcopyright{acmcopyright}\acmConference[WWW '22 Companion]{Companion Proceedings of the Web Conference 2022}{April 25--29, 2022}{Virtual Event, Lyon, France}
\acmBooktitle{Companion Proceedings of the Web Conference 2022 (WWW '22 Companion), April 25--29, 2022, Virtual Event, Lyon, France}
\acmPrice{15.00}
\acmDOI{10.1145/3487553.3524628}
\acmISBN{978-1-4503-9130-6/22/04}

\title{Analysis of Arbitrary Content on Blockchain-Based Systems using BigQuery}

\author{Marcel Gregoriadis}
\email{marcel.gregoriadis@hu-berlin.de}
\affiliation{%
  \institution{Humboldt University of Berlin}
  \state{Berlin}
  \country{Germany}
}

\author{Robert Muth}
\email{muth@tu-berlin.de}
\affiliation{%
  \institution{Technical University of Berlin}
  \state{Berlin}
  \country{Germany}
}

\author{Martin Florian}
\email{martin.florian@hu-berlin.de}
\affiliation{%
  \institution{Weizenbaum Institute}
  \state{Berlin}
  \country{Germany}
}

\begin{abstract}
Blockchain-based systems have gained immense popularity as enablers of
independent asset transfers and smart contract functionality.
They have also, since as early as the first Bitcoin blocks,
been used for storing arbitrary contents such as texts and images.
On-chain data storage functionality is useful for a variety of legitimate use cases.
It does, however, also pose a systematic risk.
If abused, for example by posting illegal contents on a public blockchain,
data storage functionality can lead to legal consequences for operators and users that need to store and distribute the blockchain,
thereby threatening the operational availability of entire blockchain ecosystems.
In this paper,
we develop and apply a cloud-based approach for quickly discovering and classifying content on public blockchains.
Our method can be adapted to different blockchain systems and offers insights into content-related usage patterns and potential cases of abuse.
We apply our method on the two most prominent public blockchain systems---Bitcoin and Ethereum---and discuss our results.
To the best of our knowledge,
the presented study is the first to systematically analyze non-financial content stored on the Ethereum blockchain and the first to present a side-by-side comparison between different blockchains in terms of the quality and quantity of stored data.

\end{abstract}



\maketitle

\section{Introduction}
Since the introduction of Bitcoin~\cite{bitcoin,tschorsch2016bitcoin} and Ethereum~\cite{ethereum},
blockchains
became an important part
for many financial-based services;
the most prevalent being cryptocurrencies.
Besides financial services, 
with the introduction of \emph{smart contracts}~\cite{szabo:smartContract, ethereum:yellowpaper}
it also became possible
to execute Turing-complete programs
and host \emph{decentralized apps}~(DApp)
on the blockchain.
On the one hand,
the general ability to insert arbitrary data into transactions
made it possible to exploit blockchains as a distributed storage medium.
On the other hand,
the ability to insert arbitrary data is a desired feature required by various DApps like
governance platforms~\cite{slockit:dao}, digital asset marketplaces~\cite{ante2021non},
and social networks~\cite{pfeiffer1232020blockchain}.

By allowing arbitrary data in transactions,
objectionable content can find its way onto the public ledger \cite{matzutt2018thwarting}.
Considering that transactions on the blockchain can be sent anonymously,
the potential for criminal intentions becomes evident.
In the case of actual illicit content,
such as
copyright violations,
personality rights violations,
or incitement to hatred,
the possession of the blockchain could become prosecuted in some jurisdictions.
This is not just a theoretical scenario:
In an analysis of the Bitcoin blockchain, \citeauthor{matzutt2018quantitative} were even able to detect,
among other things,
links to child pornography~\cite{matzutt2018quantitative}.
Since blockchain clients have to
maintain a full copy of the blockchain
for its operation,
the capability to insert arbitrary data poses a serious threat to node operators in public blockchain-based systems,
and in effect also to the resilience of the systems themselves.

In this paper,
we present and discuss the results of a quantitative and qualitative analysis of arbitrary content on the Bitcoin and Ethereum blockchains.
To this end,
we also introduce the cloud-based analysis pipeline we developed,
which uses Google BigQuery\footnote{\url{https://cloud.google.com/bigquery}}
as data source and SQL-based analysis toolkit.

We focused on Bitcoin and Ethereum which we consider to be the most relevant blockchains (with a total market capitalization of $\approx 1$~trillion USD, as of February 2nd, 2022~\cite{coinmarketcap}).
Taking into account that cryptocurrencies
to this date remain
the predominant application of blockchain technology,
and then the high combined market share of Bitcoin and Ethereum of 59.6\,\% (of which not even all are blockchain-based),
we assess our results to be representative for the emergence of arbitrary content in public blockchains.

We stress that it is not our primary goal to identify every single occurrence of arbitrary content.
Rather, our contributions can be summarized as the following:

\begin{itemize}
  \item We present and evaluate BigQuery as a cloud-based approach to analyze blockchain data.
  \item We present novel data analysis methods for the Ethereum blockchain.
  \item Using our reusable data discovery framework,
    we obtain quantitative and qualitative insights on the (non-financial) data stored on Bitcoin and Ethereum for the time frame up to January 2022,
    allowing comparisons between usage patterns
    and the general discussion of trends.
\end{itemize}

The remainder of this paper is structured as follows: 
In Section~\ref{relatedWork} we discuss related work,
also pointing out how our work compares to previous studies.
Section~\ref{sec:method} gives a general overview of our method;
foremost how we extract and filter the data that we want to examine.
Our main part, 
which is structured in Section~\ref{sec:text} and~\ref{sec:files},
specifies how we detect text and file contents, 
and gives a quantitative and qualitative overview on the results.
Section~\ref{sec:discussion} continues with an in-depth discussion of the results
and an evaluation of our methodology.
Finally, in Section~\ref{sec:conclusion}
we summarize the main conclusions of our study.

\section{Related Work}\label{relatedWork}
Since the beginning of blockchains, it has been common practice to include non-financial, arbitrary data,
as for example, in the Bitcoin genesis block~\cite{raskin2018digital}.
Non-financial data storage on blockchains enables interesting services such as
name services\footnote{\url{https://namecoin.info}},
timestamping\footnote{\url{https://opentimestamps.org}},
non-equivocation logging~\cite{tomescu2017catena},
and the management of arbitrary forms of ownership,
for example, in form of non-fungible tokens~(NFTs).
An uncensorable public data storage service,
as provided by Bitcoin and Ethereum,
can also be abused,
however.

In~\cite{matzutt2018quantitative}, \citeauthor{matzutt2018quantitative}
investigated the prevalence of non-financial data on the Bitcoin blockchain as of August 2017,
uncovering, among others, hundreds of links to child pornographic material.
The authors identified different approaches for injecting chunks of data into the Bitcoin blockchain.
On a low and general level,
these are \texttt{OP\_RETURN} output scripts, non-standard transactions, coinbase input scripts,
and pay-to-script-hash input scripts.
The authors scanned the Bitcoin blockchain for instances in which these methods have been used,
arriving at a quantitative estimate of their popularity.
To also account for data hidden in standard payment transactions,
\citeauthor{matzutt2018quantitative} additionally filtered out all transaction outputs
that contain $\geq 90 \%$ printable ASCII characters in their mutable, \ie, non-opcode part.
In addition to these low-level content detectors,
the authors also implemented methods tailored towards detecting data inserted through content insertion services such as \emph{CryptoGraffiti}\footnote{
  \url{https://cryptograffiti.info}
}.
Combing all content detection methods,
the authors concluded that \SI{1.4}{\percent} of Bitcoin transactions at the time of the study contained non-financial data.
In total,
\citeauthor{matzutt2018quantitative} were able to decode over 1,600 viewable files,
containing data such as images, text messages and source code.

Compared to the work in~\cite{matzutt2018quantitative},
we constrain our analysis to data inserted through the more general, low-level insertion techniques.
By disregarding the more specialized detection mechanisms proposed by \citeauthor{matzutt2018quantitative},
we undoubtedly miss some of their findings.
Due to the principally unbounded number of ways in which non-financial data can be encoded on the Bitcoin blockchain,
even the approach in \cite{matzutt2018quantitative} cannot be assumed to produce an exhaustive list of findings.
It must also be pointed out that any hand-tailored detectors need to be updated continuously as the landscape of content insertion services changes.
CryptoGraffiti has switched to storing data only on the \emph{Bitcoin Cash} blockchain, for example,
and is therefore not a relevant content insertion service for Bitcoin anymore.
Our cloud-based analysis toolkit,
which we have released as open source,
can flexibly be extended with hand-tailored content detectors.

Showcasing the flexibility of our cloud-based approach,
we also apply our analysis methods and tooling to uncover non-financial data stored on the Ethereum~\cite{ethereum} blockchain.
To the best of our knowledge, we are the first to conduct a systematic study into the quantity and quality of data stored on Ethereum.
Our analysis approach and tooling thereby enable insights related to the two largest blockchain networks by market capitalization of the underlying cryptocurrency.
In addition to being open source, and likely also in contrast to previously developed content detection tools,
our implementation leverages public cloud infrastructure and can thereby be used largely independently of locally available computing resources.

Various approaches have been developed for dealing with the dangers of arbitrary content insertion on public blockchains.
They can be grouped into categories as follows: avoiding
the inclusion of unwanted data~\cite{matzutt2018thwarting},
allowing the modification (and erasure) of past blockchain state~\cite{ateniese2017redactable, puddu2017muchain, deuber2019redactable, thyagarajan2021reparo},
and local pruning~\cite{florian2019erasing}.
In this paper, we focus on developing tools for determining the severity of the original problem and whether the implementation of additional protection approaches is necessary.

\section{Method}
\label{sec:method}

Our content detection and analysis methodology consists of a data extraction step and several optional post-processing and data analysis steps.
While data extraction operates on the entirety of the investigated blockchain,
post-processing and data analysis need to be performed only on the result of the first step,
\ie, on a significantly smaller amount of data.
In the following, we introduce the overall characteristics of our data extraction methodology,
the clearly more challenging part of our pipeline.
We will introduce further details on specific queries and post-processing steps in the subsequent sections
(\ref{sec:text} and \ref{sec:files}),
where we also present concrete quantitative and qualitative findings for the Bitcoin and Ethereum blockchains.
As a complement to this paper,
we release all tools and queries we developed for performing the discussed analyses as open source%
\footnote{\url{https://github.com/mg98/arbitrary-data-on-blockchains}}
,
enabling the easy reproduction of our results and the extension of our study to new areas of interest.

\subsection{Data Extraction using BigQuery}
\label{sub:data_extraction_via_bigquery}

As a distinguishing feature of our approach,
we leverage the Google Cloud BigQuery service for accessing and pre-filtering blockchain data.
BigQuery provides constantly updated datasets
for a wide range of popular public blockchains.
All data is stored in table-structured databases
which are SQL-queryable.
In BigQuery, transaction data (such as smart contracts or scripts) is provided as hexadecimal strings
which enable convenient SQL accessibility,
such as the \texttt{LIKE}-operator (for string comparisons with wildcards) 
and functions around regular expressions.

The main benefits of using the cloud service BigQuery for blockchain analysis is
that there are no (expensive) hardware requirements
and queries can be developed relatively easy.
Leveraging the provided cloud infrastructure allows to run complex queries on large datasets\footnote{At time of writing, the Bitcoin and Ethereum blockchains had a size of, respectively, more than \SI{380}{\giga\byte} and more than \SI{4}{\tera\byte} (for the full blockchain archive).}
significantly faster than with commonly available on-premise hardware.
The short response times
render the analysis process interactive
and allow adjusting queries without much effort.
Consequently, our methodology can readily be extended to support so-far overlooked data insertion techniques that might be of interest.

\subsection{Content Detection and Limitations}%
\label{sub:detecting_arbitrary_content}

For this paper,
we implemented multiple content detectors for the Bitcoin and Ethereum blockchains.
Our detectors are based on observations from practice as well as previous work such as~\cite{matzutt2018quantitative}.
They discover various texts (including URLs) and files, also reproducing previous data discoveries.
Still, our detectors clearly do not enable an \emph{exhaustive} view over \emph{all} content stored on the Bitcoin or Ethereum blockchains.
For both, Bitcoin and Ethereum, a wide range of data insertion methods is conceivable.
Many data insertion methods result in data that is only discoverable using fine-grained queries, if at all.
Encrypted data, for example, is difficult to distinguish from random noise if it does not come with identifying meta attributes (\eg, PGP headers).

In the remainder of this section,
we give an overview over our data extraction methods for Bitcoin and Ethereum.
We give more details on specific queries and post-processing steps in the subsequent sections that also describe our empirical findings.

\subsection{Content on Bitcoin}
\label{sub:method:bitcoin}

Based on the content detectors used in \cite{matzutt2018quantitative} and our own observations about commonly used approaches for inserting arbitrary content into the Bitcoin blockchain,
our content detectors for Bitcoin focus on scanning the \textit{outputs} and \textit{inputs} of transactions,
including the inputs to \textit{coinbase} transactions.

In BigQuery, the data on transaction inputs and outputs is organized in respectively named views.
Each record in the view maps to a transaction or block.
Inputs and outputs play a role outside of serving as a data store and must therefore include specific byte sequences to ensure their validity.
Therefore, the range of bytes that can be freely used for encoding arbitrary data is limited.
For our Bitcoin analysis, we analyze the concatenated bytes of these \emph{mutable values}.
From the perspective of our queries,
the investigated mutable values are the hexadecimal strings in the \textit{script\_asm} field on input and output records.
In other words, we leverage the fact that opcode bytes are made visible in the view provided by BigQuery,
and discard them for the further analysis.
In order to analyze coinbase transactions,
we contemplate the data within the \textit{coinbase\_param} field of the \textit{blocks} table provided by BigQuery.

We further classify our results by the insertion method identified. Those comprise insertions via:
\begin{itemize}
        \item \emph{Standard outputs}~(P2X),
	such as pay-to-public-key~(P2PK), 
	pay-to-public-key-hash~(P2PKH), 
	and pay-to-multi-signature (P2MS) outputs.
        \item \emph{Standard inputs}, most prominently P2PKH inputs.
        \item \emph{\texttt{OP\_RETURN} outputs}, 
	which were introduced specifically for the purpose of including arbitrary data.
        \item \emph{Non-standard inputs and outputs}, excluding \texttt{OP\_RETURN} outputs.
        \item \emph{Coinbase inputs}.
\end{itemize}

\subsection{Content on Ethereum}
\label{sub:method:ethereum}

Our Ethereum analysis is mostly limited to 
the \emph{input} field of the \emph{transactions} table in the dataset.
In the Ethereum protocol,
this field is used to deploy or call a smart contract.
Due to the variability of length and content of this field,
it has been considered the most viable option for intentional insertions of arbitrary content.

Unlike for Bitcoin, the Ethereum data set on BigQuery does not provide a field in which bytecode instructions are explicitly labeled as such.
The reason is likely the fact that the Ethereum bytecode instruction set is significantly more complex.
For the same reason, detecting bytecode instructions via a (SQL-)query is infeasible.
We therefore apply our queries on the complete and unfiltered input fields of transactions,
potentially misidentifying bytecodes as parts of stored content.
Also unlike for Bitcoin,
we ignored content (usually text) stored in coinbase transactions.
Preliminary experiments demonstrated that, as a whole,
the content stored in coinbase transactions has both an immense volume and a very low variance---%
the vast majority of coinbase transactions simply advertise the mining pool that is responsible for the current block.

\section{Text Analysis}
\label{sec:text}
 
In the following, we 
give more details on our text detection methods 
and present quantitative and qualitative text discovery results for the Bitcoin and Ethereum blockchains.
All presented results, in this and the subsequent sections,
pertain to the state of the respective blockchains on January 29th, 2022.

\subsection{Text Classification}
\label{sub:detecting_texts}

For out text analysis, we
developed a regular expression pattern to match
all kind of combinations of \mbox{UTF-8} characters
and their hexadecimal representation, respectively.
In a highly simplified manner that ignores technical steps 
such as the subtraction of opcodes,
the relevant queries for Bitcoin can be described as follows:

\begin{verbatim}
From each transaction, select
  the concatenation of all standard output scripts
    if at least 90% of the bytes in the transaction
    represent printable UTF-8 characters,
  the concatenation of all non-standard input scripts
    if 100% of the bytes in the resulting value
    represent printable UTF-8 characters,
  the concatenation of all OP_RETURN output scripts
    if 100% of the bytes after each OP_RETURN opcode
    represent printable UTF-8 characters, and
  a concatenation of all non-standard output scripts
    which are not OP_RETURN outputs
    if 100% of the bytes in the resulting value
    represent printable UTF-8 characters.
Also select the coinbase input of each block
  if 100% of its bytes
  represent printable UTF-8 characters.
\end{verbatim}

In a similar syntax, the queries for Ethereum can be written as:
\begin{verbatim}
Select the input field of each transaction
  if 100% of its bytes
  represent printable UTF-8 characters.
\end{verbatim}

As a further analysis step,
we extract the matched strings and classify them based on the occurrence of common textual data and formats.
More specifically, we classify found text blocks based on the following categories:

\begin{itemize}
  \item {\bf Strings:} Sequence of printable characters which does not contain any white spaces.
  \item {\bf Texts:} Sequence of printable characters which contains at least one white space (\ie, multiple words).
  \item {\bf Contain URL:} Contains a string that matches a URL pattern\footnotemark{}.
  \item {\bf Contain Email Address:} Contains a string that matches a generous pattern of a typical email address.
  \item {\bf Contain JSON:} Contains a string that can be decoded as a JSON object; our method does not detect simple array outputs or empty objects, \ie, ``\{\}''.
  \item {\bf Contain PGP:} Text contains a string that is enclosed with a PGP header.
  \item {\bf Contain HTML/XML:} Text contains a sequence that follows the semantics of HTML/XML (with a beginning and closing tag).
  \item {\bf Contain Data URL:} Text contains a data URL (URI scheme containing a Base64-encoded version of a file that is used to display files in-line in web pages).
\end{itemize}

As a result, our analysis returned \textbf{763,035}
corresponding transactions for Bitcoin and \textbf{1,916,836} for Ethereum.
Table~\ref{table:textType} shows the quantitative results by classification.
We point out that the majority of all text messages on Bitcoin have been embedded using \texttt{OP\_RETURN}~(78.3\,\%)
or through a coinbase transaction~(21.2\,\%).

\begin{table}[h]
\begin{center}
\begin{tabularx}{\columnwidth}{Xrrr}
\toprule
                 & \multicolumn{3}{c}{\bf Occurrences} \\
                 \cline{2-4}\\[-1em]
\bf Textual Type & \bf Total & \bf \, Bitcoin & \bf \, Ethereum \\
\midrule
Strings & 632,547 & 94,812 & 537,735 \\ 
Texts & 2,047,324 & 668,223 & 1,379,101 \\
Contain JSON & 51,128 & 2,065 & 43,063 \\
Contain HEX & 92,263 & 3,716 & 88,547 \\
Contain Email Address & 1,008 & 39 & 969 \\
Contain URL\footnotemark[\value{footnote}] & 7,435 & 4,341 & 3,094 \\
Contain PGP & 325 & 28 & 297 \\
Contain HTML/XML & 346 & 202 & 144 \\
Contain Data URL & 11 & 0 & 11 \\
\bottomrule
\end{tabularx}
\end{center}
\caption{Quantitative analysis of textual type of content.}
\label{table:textType}
\end{table}

\footnotetext{
  The analysis here operates on the results of our text detector and thereby misses transactions that are mainly non-text,
  such as smart contract invocations.
  In Section~\ref{sub:url_analysis} we discuss detecting URLs in arbitrary contexts.
}

\subsection{Frequency Analysis and Popular Strings}%
\label{sub:popular_texts}

\begin{figure}[!t]
  \centering
\begin{tikzpicture}

\definecolor{color0}{rgb}{0.12156862745098,0.466666666666667,0.705882352941177}
\definecolor{color1}{rgb}{1,0.498039215686275,0.0549019607843137}

\begin{axis}[
date coordinates in=x,
xticklabel=\year,
tick align=outside,
tick pos=left,
ytick={0, 20000, 40000, 60000, 80000, 100000},
yticklabels={0, 20\,k, 40\,k, 60\,k, 80\,k, 100\,k},
scaled y ticks = false,
x grid style={white!69.0196078431373!black},
xmin=2008-05-11 14:24, xmax=2022-01-29 20:00,
xtick style={color=black},
y grid style={white!69.0196078431373!black},
ymin=-4489.45, ymax=94872.45,
ytick style={color=black},
xlabel={Time},
ylabel={Transactions per month},
height=.5\columnwidth,
width=\columnwidth,
legend style={at={(0.03,0.8)},anchor=west},
legend cell align={left}
]
\addplot [semithick, color0]
table [header=false,col sep=comma] {%
2009-01-04 00:00,2544
2009-02-04 00:00,3380
2009-03-04 00:00,3466
2009-04-04 00:00,3442
2009-05-04 00:00,3383
2009-06-04 00:00,2236
2009-07-04 00:00,1925
2009-08-04 00:00,1564
2009-09-04 00:00,2159
2009-10-04 00:00,2126
2009-11-04 00:00,2217
2009-12-04 00:00,4048
2010-01-04 00:00,5004
2010-02-04 00:00,5603
2010-03-04 00:00,5201
2010-04-04 00:00,5578
2010-05-04 00:00,4940
2010-06-04 00:00,4746
2010-07-04 00:00,7814
2010-08-04 00:00,5465
2010-09-04 00:00,5374
2010-10-04 00:00,5219
2010-11-04 00:00,5741
2010-12-04 00:00,5599
2011-01-04 00:00,4980
2011-02-04 00:00,5153
2011-03-04 00:00,4541
2011-04-04 00:00,4845
2011-05-04 00:00,6587
2011-06-04 00:00,6063
2011-07-04 00:00,4050
2011-08-04 00:00,3614
2011-09-04 00:00,3259
2011-10-04 00:00,2304
2011-11-04 00:00,2338
2011-12-04 00:00,2463
2012-01-04 00:00,2553
2012-02-04 00:00,2379
2012-03-04 00:00,1290
2012-04-04 00:00,373
2012-05-04 00:00,362
2012-06-04 00:00,342
2012-07-04 00:00,286
2012-08-04 00:00,294
2012-09-04 00:00,254
2012-10-04 00:00,273
2012-11-04 00:00,238
2012-12-04 00:00,253
2013-01-04 00:00,256
2013-02-04 00:00,243
2013-03-04 00:00,222
2013-04-04 00:00,217
2013-05-04 00:00,160
2013-06-04 00:00,98
2013-07-04 00:00,88
2013-08-04 00:00,73
2013-09-04 00:00,66
2013-10-04 00:00,146
2013-11-04 00:00,106
2013-12-04 00:00,287
2014-01-04 00:00,116
2014-02-04 00:00,104
2014-03-04 00:00,50
2014-04-04 00:00,27
2014-05-04 00:00,56
2014-06-04 00:00,37
2014-07-04 00:00,129
2014-08-04 00:00,218
2014-09-04 00:00,297
2014-10-04 00:00,203
2014-11-04 00:00,312
2014-12-04 00:00,433
2015-01-04 00:00,516
2015-02-04 00:00,446
2015-03-04 00:00,1187
2015-04-04 00:00,951
2015-05-04 00:00,1162
2015-06-04 00:00,1680
2015-07-04 00:00,2389
2015-08-04 00:00,1983
2015-09-04 00:00,40604
2015-10-04 00:00,8774
2015-11-04 00:00,6170
2015-12-04 00:00,5221
2016-01-04 00:00,9977
2016-02-04 00:00,12002
2016-03-04 00:00,21871
2016-04-04 00:00,11394
2016-05-04 00:00,7416
2016-06-04 00:00,8455
2016-07-04 00:00,18139
2016-08-04 00:00,6813
2016-09-04 00:00,4889
2016-10-04 00:00,5247
2016-11-04 00:00,5918
2016-12-04 00:00,5054
2017-01-04 00:00,5293
2017-02-04 00:00,12102
2017-03-04 00:00,24057
2017-04-04 00:00,39844
2017-05-04 00:00,14486
2017-06-04 00:00,5940
2017-07-04 00:00,3015
2017-08-04 00:00,8810
2017-09-04 00:00,4845
2017-10-04 00:00,12617
2017-11-04 00:00,5722
2017-12-04 00:00,5180
2018-01-04 00:00,6835
2018-02-04 00:00,969
2018-03-04 00:00,805
2018-04-04 00:00,1714
2018-05-04 00:00,1174
2018-06-04 00:00,1205
2018-07-04 00:00,1082
2018-08-04 00:00,1092
2018-09-04 00:00,1099
2018-10-04 00:00,3127
2018-11-04 00:00,6932
2018-12-04 00:00,6489
2019-01-04 00:00,10253
2019-02-04 00:00,10531
2019-03-04 00:00,25580
2019-04-04 00:00,7920
2019-05-04 00:00,7897
2019-06-04 00:00,3543
2019-07-04 00:00,12343
2019-08-04 00:00,3472
2019-09-04 00:00,2098
2019-10-04 00:00,2219
2019-11-04 00:00,2001
2019-12-04 00:00,2479
2020-01-04 00:00,1896
2020-02-04 00:00,1250
2020-03-04 00:00,7967
2020-04-04 00:00,4971
2020-05-04 00:00,1298
2020-06-04 00:00,1913
2020-07-04 00:00,1517
2020-08-04 00:00,1264
2020-09-04 00:00,1663
2020-10-04 00:00,1065
2020-11-04 00:00,1309
2020-12-04 00:00,3324
2021-01-04 00:00,3665
2021-02-04 00:00,2533
2021-03-04 00:00,6674
2021-04-04 00:00,6524
2021-05-04 00:00,16647
2021-06-04 00:00,11534
2021-07-04 00:00,7374
2021-08-04 00:00,5891
2021-09-04 00:00,5186
2021-10-04 00:00,13471
2021-11-04 00:00,12758
2021-12-04 00:00,13556
2022-01-04 00:00,12945
};
\addplot [semithick, color1]
table [header=false,col sep=comma] {%
2015-08-04 00:00,3144
2015-09-04 00:00,746
2015-10-04 00:00,680
2015-11-04 00:00,353
2015-12-04 00:00,679
2016-01-04 00:00,1843
2016-02-04 00:00,1652
2016-03-04 00:00,1303
2016-04-04 00:00,580
2016-05-04 00:00,1635
2016-06-04 00:00,32478
2016-07-04 00:00,20072
2016-08-04 00:00,3142
2016-09-04 00:00,1194
2016-10-04 00:00,159
2016-11-04 00:00,395
2016-12-04 00:00,137
2017-01-04 00:00,492
2017-02-04 00:00,409
2017-03-04 00:00,36094
2017-04-04 00:00,10378
2017-05-04 00:00,5366
2017-06-04 00:00,55142
2017-07-04 00:00,39801
2017-08-04 00:00,45535
2017-09-04 00:00,28365
2017-10-04 00:00,12093
2017-11-04 00:00,17908
2017-12-04 00:00,38236
2018-01-04 00:00,36269
2018-02-04 00:00,16293
2018-03-04 00:00,23067
2018-04-04 00:00,18782
2018-05-04 00:00,34192
2018-06-04 00:00,65322
2018-07-04 00:00,36636
2018-08-04 00:00,90356
2018-09-04 00:00,25894
2018-10-04 00:00,23727
2018-11-04 00:00,27555
2018-12-04 00:00,42502
2019-01-04 00:00,45745
2019-02-04 00:00,21265
2019-03-04 00:00,33621
2019-04-04 00:00,41193
2019-05-04 00:00,49184
2019-06-04 00:00,26845
2019-07-04 00:00,29953
2019-08-04 00:00,31423
2019-09-04 00:00,28735
2019-10-04 00:00,35609
2019-11-04 00:00,32936
2019-12-04 00:00,53072
2020-01-04 00:00,32591
2020-02-04 00:00,19526
2020-03-04 00:00,25911
2020-04-04 00:00,27650
2020-05-04 00:00,38511
2020-06-04 00:00,36618
2020-07-04 00:00,33302
2020-08-04 00:00,27384
2020-09-04 00:00,23537
2020-10-04 00:00,23368
2020-11-04 00:00,25379
2020-12-04 00:00,21115
2021-01-04 00:00,32502
2021-02-04 00:00,26801
2021-03-04 00:00,22655
2021-04-04 00:00,28438
2021-05-04 00:00,24443
2021-06-04 00:00,23397
2021-07-04 00:00,21735
2021-08-04 00:00,20566
2021-09-04 00:00,27049
2021-10-04 00:00,25954
2021-11-04 00:00,24883
2021-12-04 00:00,27190
2022-01-04 00:00,46148
};
\addlegendentry{Bitcoin}
\addlegendentry{Ethereum}
\end{axis}

\end{tikzpicture}
\vspace*{-2.25em}
  \caption{Frequency of text transactions over time.}
\label{fig:textFreq}
\end{figure}
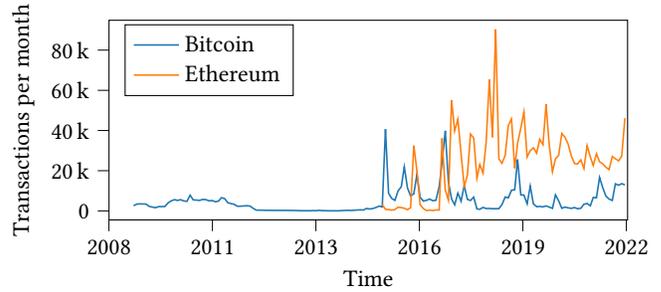

\begin{figure}[!t]
  \centering
  \input{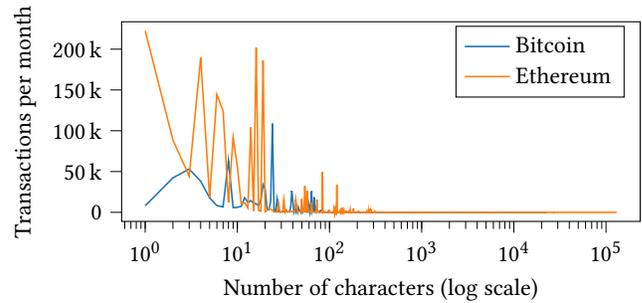}
  \caption{Frequency of text lengths.}
\label{fig:textLength}
\end{figure}

In the following,
we analyzed the corresponding transactions and 
counted the occurrences of extracted texts.
Figure~\ref{fig:textFreq} shows the frequency of text messages in transactions over time, for both blockchains.
In Figure~\ref{fig:textLength}, we depict the frequency of texts by length.
As can be seen, most of the texts are short.
The closer inspection of a sample of the results made it obvious that a lot of messages appear several times
or have a similar structure.
We have listed the most common embedded texts in Table~\ref{table:textRanking:bitcoin} and~\ref{table:textRanking:ethereum}.
Further investigation has shown that not only those texts
but also texts sharing the same structure caused the peaks in Figure~\ref{fig:textLength}.
We have also observed that those transactions would often be received by the same recipient,
\eg, Kraken\footnote{\url{https://kraken.com}}.

\begin{table}[h]
\begin{center}
\begin{tabularx}{\columnwidth}{lXr}
\toprule
\bf \# & \bf Text & \bf \hspace{-3em}Occurrences \\
\midrule
1 & Bitzlato & 60,625 \\
2 & ASCRIBESPOOL01PIECE & 28,986 \\
3 & 503: Bitcoin over capacity! & 13,902 \\
4 & ``2265861855@qq.com''\} & 12,880 \\
5 & We'll buy your Bitcoins. sell.buy.bitcoin@\dots & 10,790 \\
6 & ``2265861855@QQ.COM''\} & 9,800 \\
7 & WWW.BTCKEY.ORG Bitcoin wallet recovery\dots & 5,530 \\
8 & ASCRIBESPOOL01FUEL & 5,125 \\
9 & Bitcoin: A Peer-to-Peer Electronic Cash System & 4,843 \\
10 & ASCRIBESPOOL01EDITIONS1 & 4,270 \\
\bottomrule
\end{tabularx}
\end{center}
\caption{Top 10 most frequently embedded texts found on Bitcoin (\#5 and \#7 have been truncated).}
\label{table:textRanking:bitcoin}
\end{table}

\begin{table}[h]
\begin{center}
\begin{tabularx}{\columnwidth}{lXr}
\toprule
\bf \# & \bf Text & \bf Occurrences \\
\midrule
1 &   & 196,211 \\
2 & BFX\_REFILL\_SWEEP & 179,820 \\
3 & hotwallet drain fee & 177,454 \\
4 & Ignore & 126,140 \\
5 & imtoken & 96,807 \\
6 & coinbenerefuel & 95,385 \\
7 & undefined & 61,226 \\
8 & fz\/X & 55,924 \\
9 & FA\%\} & 54,684 \\
10 & cs & 48,079 \\
\bottomrule
\end{tabularx}
\end{center}
\caption{Top 10 most frequently embedded texts found on Ethereum.}
\label{table:textRanking:ethereum}
\end{table}

\vspace*{-5mm}

\subsection{Qualitative Observations}%
\label{sub:qualitative_observations}

Individual content analysis has shown many messages, 
even bidirectional conversations, 
to when funds were accidentally sent to the wrong address or when funds got stolen from a user. 
The victim would then try to ask or negotiate to get his or her funds back. 
Presumably to save on transaction fees, 
some longer messages were shared through the URL of an online service like \emph{PasteBin}\footnote{\url{https://pastebin.com}}.
We have highlighted a selection of text messages in Appendix~\ref{appendix}.

We also decoded the 11 data URLs that were found to image type files (JPEG, PNG, GIF).
One of those was a provocative image showing the Chinese president Xi Jinping as Winnie the Pooh.

\subsection{Analysis of URLs}
\label{sub:url_analysis}

URLs persisted on the blockchain can point to more rich resources in the internet, 
such as images, videos, long texts, and other files.
This is interesting for qualitative observations
and also in search of illegal contents.
With this in mind, we created a method to scan the Ethereum blockchain
more exhaustively for URLs.
As opposed to our basic text analysis, 
we now aim at finding URL strings 
also within smart contract calls and deployments.
To this end, we created regular expression patterns to match with HTTP and IPFS~\cite{benet2014ipfs} URLs, 
with extra attention to URLs pointing to Tor~\cite{dingledine2004tor} onion services.

In simplified notation that abstracts away technical details such as the encoding and decoding of strings,
our main URL detector logic can be written as follows:
\begin{verbatim}
Select
  all transaction input fields on Ethereum
with occurrences of
  http://*, https://*, ipfs://*
where * matches
  the longest coherent string of URL-valid characters
  and is at least 5 characters long.
Also select
  all transaction input fields on Ethereum
with occurrences of
  *.onion
where * matches
  the longest coherent string of alphanumeric characters
  of at least 16 characters.
\end{verbatim}
Further extraction, verification, and classification into HTTP, HTTPS, IPFS and .onion links happens in a subsequent post-processing step.
For Bitcoin, we could perform those steps directly on the result set from the text analysis.
The quantitative results can be observed from Table~\ref{table:urlQuantities}.

\begin{table}[h]
\begin{center}
\begin{tabularx}{\columnwidth}{Xrrr}
\toprule
& \multicolumn{3}{c}{\bf Occurrences} \\
\cline{2-4}\\[-1em]
\bf URL Type & \bf Total & \bf Bitcoin & \bf Ethereum \\
\midrule
HTTP & 522,313 & 4,329 & 517,985 \\
IPFS & 210,268 & 7 & 210,261 \\
Onion Service & 52 & 6 & 46 \\
\hline
\bf Sum & \bf 732,633 & \bf 4,388 & \bf 728,292 \\
\bottomrule
\end{tabularx}
\end{center}
\caption{Quantitative analysis of discovered URLs.}
\label{table:urlQuantities}
\end{table}

In addition to the quantitative analysis,
we qualitatively analyzed a sample of 100 HTTP links chosen randomly from our results for each blockchain.
Based on our manual analysis, we make the following observations about our sample:

\textbf{Bitcoin:} 19\,\% of the URLs linked to cryptocurrency-related content, 
another 11\,\% linked to social media content (Twitter, Reddit, and YouTube), 
2\,\% showed pornography, 
and 25\,\% were miscellaneous. 
43\,\% of the links were dead
\footnote{This includes unavailable websites as well as content platforms (\eg, social media sites or file sharing services) which show that the requested resource has been removed or access is restricted.}, 
however most of them were obviously relating to cryptocurrencies (judging by the URL).
Also most social media posts related to cryptocurrencies.

\textbf{Ethereum}:
14\,\% referred to cryptocurrency-related content, 
6\,\% of the URLs responded with a JSON that share characteristics of NFT metadata
(comprising keys such as ``name'', ``description'', ``image'', and an array of ``attributes''),
and 33\,\% were miscellaneous (\eg, memes).
Here as well, 43\,\% of the links were dead.

From all detected onion service URLs across both blockchains,
we were able to access only 3 unique services:
an online copy of the bible,
a mirror of the official website of the CIA,
and a website hosting various texts with ``forbidden knowledge'' (including tutorials on drug synthesis and explosives).
While the other services appeared to be unavailable,
online research suggests that some of the found URLs might have linked to child pornography content in the past.

Note that our URL detection method has two explicit limitations:
\begin{enumerate}
  \item We ignore the possibility of a single transaction carrying multiple URLs and extract (and count) only the first match.
  \item The pattern we use to find URLs will sometimes wrongly match additional bytes located directly after the end of a matched URL.
\end{enumerate}

The situation outlined in~(2) can for example happen when the URL appears inside a JSON structure and our pattern interprets the proceeding ``"\}'' as a part of the HTTP path.
It can also happen with random ASCII characters caused by general noise.
Another cause, in Ethereum, can be direct concatenation of string parameters in a smart contract call.
The following string is an actual URL matched by our program:

\begin{verbatim}
	https://file.soar.earth/d4c4540faf449a9a729edbf9e60d3
	621.jpg/previewGhttps://api.soar.earth/v1/download/d4
	c4540faf449a9a729edbf9e60d3621.jpg+POINT(115.63155412
	67395
\end{verbatim}

This example demonstrates the issue raised in~(2):
Our detectors wrongly matched the subsequent parameter and even a part of another, third parameter in a smart contract call.
Syntactically, all of this could still constitute a valid URL,
while semantically it is likely the case that they are actually two URLs hidden in the returned string.

Limitations~(1) and~(2) have a limited impact on our results, however.
While the quantitative analysis might be impacted by~(1),
manual observation of samples of our collected data suggests that transactions with multiple URLs are rare.
Regarding the impact of~(2) on qualitative analyses,
note that our qualitative analysis was performed manually on randomly sampled subset of all findings,
which allowed us to manually fix any obviously erroneous URL add-ons.

\section{Files Analysis}
\label{sec:files}
Besides text-based arbitrary data,
we also searched the Bitcoin and Ethereum blockchains for whole files.
In the following,
we show how to detect different types of files
and 
present our results on a quantitative and qualitative basis.

\subsection{Detecting Files and File Types}
\label{sub:detecting_files_and_file_types}

In order to find files on the blockchain,
we scanned all transactions 
for the occurrence of popular file type signatures.
In the analysis for both blockchains, 
findings were evaluated 
from the start of a signature to the very end of the transaction's payload.
To this end, we stripped out the opcodes from the payload for the analysis on Bitcoin,
and left out file types with very short signatures 
(\eg, GZIP or DOS executables) 
because they caused too many false positives.

In a very simplified manner,
the queries we used to retrieve our result candidates for Bitcoin
can be formulated as follows:

\begin{verbatim}
From each transaction, select
  the concatenation of all output scripts,
  the concatenation of all non-standard input scripts,
  and the concatenation of all pay-to-script-hash 
  input scripts,
with at least one occurrence of the byte sequence of
  a file signature.
Also select the coinbase input of each block
with at least one occurrence of the byte sequence of
  a file signature.
\end{verbatim}

The respective query for Ethereum can be written as:
\begin{verbatim}
Select
  all transaction input fields
with at least one occurrence of the byte sequence of
  a file signature.
Also select the coinbase input of each block
with at least one occurrence of the byte sequence of
  a file signature.
\end{verbatim}

In our analysis of Ethereum, we later distinguish our findings between two insertion methods:

\begin{itemize}
  \item \textbf{Embedded:} Transaction data represents the encoded file.
  \item \textbf{Injected:} Transaction data does not start with but ends with the encoded file. This will be considered a smart contract call where the last argument has been the file.
\end{itemize}

We also examined coinbase transactions in Ethereum but report zero findings there.
This is unsurprising given the fact that the coinbase of Ethereum blocks can hold only 32~bytes---too little space for anything but short text messages.

Since searching for longer file signatures
still leads to a high number of false positives, 
we developed post-processing scripts which 
filter files that cannot be opened with standard software.
The scripts also unveil Microsoft Word documents~(DOC) that have been identified as ZIP archives before.
In the next step, we go over the results manually and remove remaining broken files that our automated post-processing did not filter out.

\subsection{Frequency of File Types}
\label{sub:quantitative_results}
As shown in Table~\ref{table:fileTypes}, 
we focus on the most popular media file types, documents, and archives.
As a result of the post-processing,
we analyze the quantitative results for readable files found on both blockchains.
As a matter of fact, 764 of the 847 findings were images,
also including many duplicates.
The majority of images on Ethereum were injected in the context of NFT projects. 
Other images have partly been categorized, as Table~\ref{table:fileCategories} shows. 

As of Bitcoin, 
we found 77.0\,\% inserted through P2X transactions 
and 21.8\,\% inserted through P2SH input scripts.
A single occurrence was found in a non-standard output script.

\begin{table}[h]
\begin{center}

\subfloat[Results for Bitcoin.]{
	\begin{tabularx}{\columnwidth}{Xr}
	\toprule
	\bf File Type & \bf Total \\ 
	\midrule
	PNG & 38 \\
	JPEG & 42 \\
	GIF & 2 \\
	PDF & 2 \\
	ZIP & 2 \\
	7-ZIP & 1 \\
	WEBP, DOC, MP3, MP4, MOV, WAV, AVI, RAR, TAR & 0 \\
	\hline
	\bf Sum & \bf 87 \\
	\bottomrule
	\end{tabularx}
}\label{table:fileTypes:bitcoin}
\subfloat[Results for Ethereum.]{
\begin{tabularx}{\columnwidth}{Xrrr}
\toprule
\bf File Type & \bf Total & \bf Embed. & \bf Injected \\ 
\midrule
	PNG & 275 & 50 & 225 \\
	WEBP & 275 & 0 & 275 \\
	JPEG & 124 & 72 & 52 \\
	7-ZIP & 68 & 68 & 0 \\
	GIF & 8 & 6 & 2 \\
	ZIP & 4 & 4 & 0 \\
	MP3 & 3 & 1 & 2 \\
	PDF & 2 & 2 & 0 \\
	DOC & 1 & 1 & 0 \\
	MP4, MOV, WAV, AVI, RAR, TAR & 0 & 0 & 0 \\
	\hline
	\bf Sum & \bf 760 & \bf 204 & \bf 556 \\
	\bottomrule
	\end{tabularx}
}\label{table:fileTypes:ethereum}

\end{center}
\caption{Quantitative results of viewable files.}
\label{table:fileTypes}
\end{table}

\subsection{Examples of Found Images}%
\label{sub:types_of_found_images}

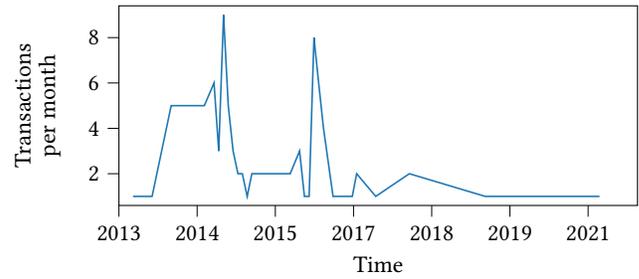
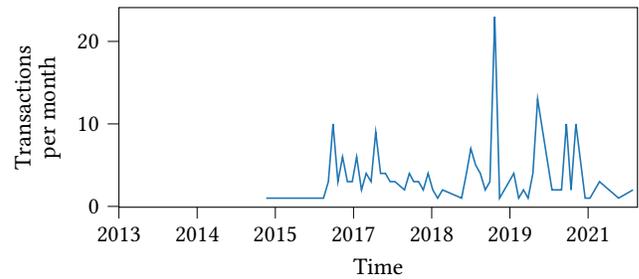
\begin{figure}[!tbp]
  \centering
  \subfloat[Results for Bitcoin.]{
\begin{tikzpicture}

\definecolor{color0}{rgb}{0.12156862745098,0.466666666666667,0.705882352941177}

\begin{axis}[
date coordinates in=x,
xticklabel=\year,
tick align=outside,
tick pos=left,
ytick={2,4,6,8},
yticklabels={\phantom{0}2, 4, 6, 8}, 
xmin=2013-01-01 00:00, xmax=2022-01-29 20:00,
xtick style={color=black},
x grid style={white!69.0196078431373!black},
ymin=0.6, ymax=9.4,
ytick style={color=black},
xlabel={Time},
ylabel={\shortstack{Transactions\\per month}},
height=.5\columnwidth,
width=\columnwidth
]
\addplot [semithick, color0]
table [header=false,col sep=comma] {%
2013-04-01 00:00,1
2013-08-01 00:00,1
2013-12-01 00:00,5
2014-07-01 00:00,5
2014-09-01 00:00,6
2014-10-01 00:00,3
2014-11-01 00:00,9
2014-12-01 00:00,5
2015-01-01 00:00,3
2015-02-01 00:00,2
2015-03-01 00:00,2
2015-04-01 00:00,1
2015-05-01 00:00,2
2016-01-01 00:00,2
2016-03-01 00:00,3
2016-04-01 00:00,1
2016-05-01 00:00,1
2016-06-01 00:00,8
2016-07-01 00:00,6
2016-08-01 00:00,4
2016-10-01 00:00,1
2016-11-01 00:00,1
2016-12-01 00:00,1
2017-01-01 00:00,1
2017-02-01 00:00,1
2017-03-01 00:00,2
2017-07-01 00:00,1
2018-02-01 00:00,2
2019-06-01 00:00,1
2019-10-01 00:00,1
2020-04-01 00:00,1
2020-10-01 00:00,1
2021-01-01 00:00,1
2021-05-01 00:00,1
2021-06-01 00:00,1
};
\end{axis}

\end{tikzpicture}\label{fig:freqFiles:bitcoin}}
  \hfill
  \subfloat[Results for Ethereum (excluding files in smart contract calls).]{
\begin{tikzpicture}

\definecolor{color0}{rgb}{0.12156862745098,0.466666666666667,0.705882352941177}

\begin{axis}[
date coordinates in=x,
xticklabel=\year,
tick align=outside,
tick pos=left,
xmin=2013-01-01 00:00, xmax=2022-01-29 20:00,
xtick style={color=black},
y grid style={white!69.0196078431373!black},
ymin=-0.1, ymax=24.1,
ytick style={color=black},
xlabel={Time},
ylabel={\shortstack{Transactions\\per month}},
height=.5\columnwidth,
width=\columnwidth
]
\addplot [semithick, color0]
table [header=false,col sep=comma] {%
2015-08-01 00:00,1
2016-04-01 00:00,1
2016-05-01 00:00,1
2016-08-01 00:00,1
2016-09-01 00:00,3
2016-10-01 00:00,10
2016-11-01 00:00,3
2016-12-01 00:00,6
2017-01-01 00:00,3
2017-02-01 00:00,3
2017-03-01 00:00,6
2017-04-01 00:00,2
2017-05-01 00:00,4
2017-06-01 00:00,3
2017-07-01 00:00,9
2017-08-01 00:00,4
2017-09-01 00:00,4
2017-10-01 00:00,3
2017-11-01 00:00,3
2018-01-01 00:00,2
2018-02-01 00:00,4
2018-03-01 00:00,3
2018-04-01 00:00,3
2018-05-01 00:00,2
2018-06-01 00:00,4
2018-07-01 00:00,2
2018-08-01 00:00,1
2018-09-01 00:00,2
2019-01-01 00:00,1
2019-02-01 00:00,4
2019-03-01 00:00,7
2019-04-01 00:00,5
2019-05-01 00:00,4
2019-06-01 00:00,2
2019-07-01 00:00,3
2019-08-01 00:00,23
2019-09-01 00:00,1
2019-12-01 00:00,4
2020-01-01 00:00,1
2020-02-01 00:00,2
2020-03-01 00:00,1
2020-04-01 00:00,4
2020-05-01 00:00,13
2020-08-01 00:00,2
2020-09-01 00:00,2
2020-10-01 00:00,2
2020-11-01 00:00,10
2020-12-01 00:00,2
2021-01-01 00:00,10
2021-03-01 00:00,1
2021-04-01 00:00,1
2021-06-01 00:00,3
2021-10-01 00:00,1
2022-01-01 00:00,2
};
\end{axis}

\end{tikzpicture}\label{fig:freqFiles:ethereum}}
  \caption{Frequency of transactions with embedded files over time.}
\label{fig:freqFiles}
\end{figure}

\begin{table}[h]
\begin{center}
\begin{tabularx}{\columnwidth}{Xrrr}
\toprule
& \multicolumn{3}{c}{\bf Occurrences} \\
\cline{2-4}\\[-1em]
\bf Category  & \bf Total & \bf Bitcoin & \bf Ethereum \\
\midrule
Portraits & 18 & 6 & 17 \\
Memes & 16 & 4 & 12 \\
Crypto Related & 11 & 4 & 7 \\
Erotics & 10 & 3 & 7 \\
Family/Group Photos & 7 & 3 & 4 \\
Text as Image & 8 & 0 & 8 \\
Cats & 5 & 1 & 4 \\
Explicit Pornography & 3 & 1 & 2 \\
\bottomrule
\end{tabularx}
\end{center}
\caption{Categorization of a selection of images found in the analysis (not counting duplicates).}
\label{table:fileCategories}
\end{table}

In the context of the (pseudonymous) anonymity and irreversibility of blockchain transactions,
we noted the following selection of found images:

\begin{itemize}
  \item a swastika
  \item a photo of a birth certificate laying on a newborn baby
  \item an academic degree
  \item a screenshot of the Twitter app showing the Chinese ambassador Liu Xiaoming having liked a post containing suggestive content
  \item the president of Russia Vladimir Putin in a LGBTQ theme
  \item a meme making fun of the supreme leader of North Korea Kim Jong-un
\end{itemize}

\subsection{Discussion of Non-Image Files}%
\label{sub:non_image_files}

We also investigated files which are not images.
In sum, we found 78 non-image files
which we were able to decode.
As a result of a manual, predominantly qualitative analysis, we
highlight a number of noteworthy findings.

We found several audio files, as for example, 7~seconds of electronic music in generally poor quality.
Two other \emph{audio files} played a brief audio interference sound.
However, we assume that there are more audio files on the blockchain
which we were just not able to decode correctly.

We decoded 4~\emph{PDF documents} which basically contained:
\begin{itemize}
    \item a white page with the text ``Ethereum White Paper''
    \item the original white paper for Bitcoin
    \item a manual on Adobe Acrobat PDF files
    \item a white page
\end{itemize}

The only \emph{DOC file} we found contained 3~pages of Russian text, beginning with the headline \textit{``So, you're reading this text when I've been dead for, I think, at least a few centuries.''} (translated), signed January~23, 2019.
The document also referenced a transaction on the blockchain with a 7-ZIP archive, which
contained a text file with various links to images, videos, audio files, and other text documents.
Most of them were captured by the \emph{Internet Archive}\footnote{\url{https://web.archive.org}}
and basically contained either poetic or philosophical statements.

Even though the majority of found archives was password protected, we successfully decompressed 6~\emph{ZIP archives}:
\begin{itemize}
	\item configuration files for the music-related software \emph{Ivory}
	\item a digital certificate in DER format
    \item a static HTML website which presents itself as a proof of concept for an honors thesis at Albion College, presumably about etching content onto the blockchain
    \item a static HTML website to apparently find and play songs stored on the Ethereum blockchain using JavaScript
    \item the C\# source code for the program \emph{minicryptowallet}
    \item a static HTML website with text mostly in Russian which shares great similarity to the findings of the previously described DOC file
\end{itemize}

\section{Discussion}\label{sec:discussion}

In this section we dive deeper into the key results of our analyses
and discuss their implications.

\subsection{BigQuery and Our Method}

During the development of our analyzer,
BigQuery has proven to be a robust, easy-to-use, and flexible
platform to scrape structured data from a blockchain; 
results were returned within just a few minutes or a few hours depending on the analysis.
BigQuery supports a very powerful set of SQL functions.
Though, further methods are necessary to process and analyze the results;
this especially accounts for the file results.
We also encountered limitations with subqueries and regular expressions.
BigQuery restricts a regular expression pattern to have no more than one capturing group
which make effective pattern matching, \eg, with URLs, difficult.
It nevertheless enables great first filtering of the data 
that can afterwards be validated and processed by other programs.

\subsection{Quality of Results}

For the following discussion of our findings,
we point out that our method did not find (or evaluate) every arbitrary content on Bitcoin or Ethereum.
Additionally, many of our specific findings are likely to be corrupted,
in the sense that they contain additional artifact or represent incomplete blobs of content.
One reason lies in the fact that we never combine data spread across multiple transactions.
Especially bigger files are likely to be spread over multiple transactions, however.
This becomes very apparent with cut image files which we encountered with a significant number of findings.
We observe the same things for text results in Bitcoin,
where we sometimes saw fragments of JSON objects.
Additionally,
we suspect that many of the very short texts we detected,
especially on Bitcoin,
might be false positives.
Those are naturally very hard to distinguish from noise.
We compared our results to the numbers of detected transactions per low-level insertion methods in~\cite{matzutt2018quantitative}
and confirm similar results with our method (for the respective time frame).
Quantitatively, we were also able to replicate the file results (disregarding files found by the service detectors).

\subsection{How Content Storage Possibilities are Used}

The text analysis has shown
that the input bytes field in transactions was mainly used
as a way to store non-financial information using the Ethereum protocol,
but we also identified a significant amount of organizations that utilize this field to establish their own protocols (\eg, to transfer JSON objects).

The files analysis has shown a number of results
that originate from people just being enthusiastic about the possibility to persist arbitrary content in the blockchain.
Our qualitative observation revealed
time capsuled contents, with the intention to be discovered much later in the future
or to persist them irrevocably (\eg, official certificates).

Both analyses have shown a lot of duplicate content 
and specifically for the file results it was evident that few individuals are responsible for a large parts of insertions.
A large part of the file contents and even a lot of text messages 
(based on the review of a sample of results)
are related to cryptocurrencies.
It is even more obvious in our qualitative evaluation of a sample of the found URLs,
where cryptocurrency-related content constitutes the largest category.
The popularity of cryptocurrency-related content does not come as a surprise 
since the insertion of arbitrary content (outside of smart contracts)
requires greater technical engagement with the system
and methods that are not as accessible to regular users.

In almost every quantitative assessment of results, 
Ethereum scored significantly higher than Bitcoin, 
even though Ethereum got introduced eight years later than Bitcoin.
We reason this by arbitrary content insertions 
being cheaper and generally more convenient 
on Ethereum than on Bitcoin.
Up until mid-2020, the transaction fees in Ethereum used to be consistently lower,
most of the time only a fraction of the fees charged for transactions on Bitcoin \cite{txFeesBtc,txFeesEth}.
A similar correlation can be inferred from Figure~\ref{fig:textFreq}.
Furthermore, transactions in Ethereum are far more flexible
and it has not been until 2014 that Bitcoin introduced
a more accessible solution to this with \texttt{OP\_RETURN}~\cite{opreturn}.
But even with this solution, 
the space in a single transaction is relatively small
compared to Ethereum's input field 
which was designed to carry the byte code for complex smart contract logic.
This can become a constraint when uploading larger files,
forcing users to split the bytes of a file to multiple transactions.
This ultimately makes Ethereum the more attractive choice,
especially for file insertions.

\subsection{Legal Assessment of Results}

In our efforts to make a statement on the legal risks resulting from injected contents,
we have paid special attention to URLs detected in our text analysis.
While we were not able to review every single URL, due to the high quantity,
we were able to link at least one finding (in each blockchain) to former backup URLs to child pornography websites.
The file analysis has also led to some sensitive or offensive contents (\eg, a swastika and explicit pornography).
We found one image that depicts, in low resolution, mild nudity of a woman that could potentially be a minor.
All in all, very few of our findings come close to being considered \enquote{illegal} by Western countries' standards.
We also did not find any obvious copyright violations.
We did, however, find images offending leaders of
countries such as Russia, China, and North Korea,
the possession and redistribution of which could be frowned upon within the respective countries.
Likewise,
the possession and redistribution of pornographic content and arbitrary religious texts is also not universally accepted or even legal in all jurisdictions.

A remaining source of potential legal risks lies in the large amount on privacy-relevant data that is stored on both investigated blockchains,
including personal photographs and personal details such as birth dates.
To which extent these occurrences could lead to clashes with existing data protection legislation and personality rights remains an open question.
Recall that erasing data from blockchains, which can be mandated by a legal framework such as the EU's GDPR~\cite{eu2016gdpr},
poses a significant challenge~\cite{ateniese2017redactable, puddu2017muchain, florian2019erasing, deuber2019redactable, thyagarajan2021reparo}.

\section{Conclusion}
\label{sec:conclusion}

We presented a novel approach for detecting and classifying arbitrary non-financial content on public blockchains using
the Google Cloud service BigQuery,
and applied it to the Bitcoin and Ethereum blockchains.
We discovered arbitrary contents of various types;
ranging from peer-to-peer communication, 
advertisement,
philosophical messages,
to various forms of structured data, presumably triggering external logic.
Finally, we also confirm former results indicating
that the possibility to post arbitrary data
has been abused to post illegal or objectionable content,
such as images or URLs to externally hosted (\eg, via onion services) problematic content.
Furthermore, we have evaluated BigQuery as a means to perform analyses of blockchain data.
While we have identified technical limitations,
we still assess BigQuery to be a useful and appropriate base for further, more complex analyses.

\newpage
\bibliographystyle{ACM-Reference-Format}
\bibliography{references}

\newpage

\begin{appendix}
\onecolumn

\section{Selection of Findings from the Text Analysis}\label{appendix}

\vspace*{4mm}
\begin{itemize}
	\item ``EW You can now use app.remembr.io to send any kind of message on blockchain'' \\[2mm] \textit{(Found on Bitcoin. Hash: 095f90b1414ccc90d0615a447b0a82847e7aa1bb70807d82a613b48a3043ec49. Aug-31-2015 01:19:22 PM +UTC.)} \\[1mm]
	\item ``L176,Obama spent about \$65,000 of the tax-payers money flying in pizza/dogs'' \\[2mm] \textit{(Found on Bitcoin. Hash: d7c7e871f6502179761122d7fc7d0ed7fa367cd2cf38c46c8d177a119d343437. Jan-04-2017 02:11:53 PM +UTC.)} \\[1mm]
	\item ``Happy Birthday Joachim! With love from the Jolocom team!'' \\[2mm] \textit{(Found on Bitcoin. Hash: d00dab66b0ef0142dbfb51f51beac3eb33fc5dd7635539a5deccdffdf109d019. Jan-21-2018 10:35:49 AM +UTC.)} \\[1mm]
    \item ``hello, guy, can you send my 5.1558763eth back ? that is all my currency in the crypto world, that money I am ready to get married, and I thought  I could earn double this time, but now I couldn't get it back. i don't know why this happen.  If you send it to me back, i would really appreciate you!!!'' \\[2mm] \textit{(Found on Ethereum. Hash: 0x884ea1daf9888e5c1aa9d12737bc90e186eb08a6ced9fce9595fbe4878b645c0. Jun-15-2021 02:04:16 PM +UTC.)} \\[1mm]
    \item ``Madeleine, I love you so much.  My love for you is eternal like this message. Happy 6 Months!'' \\[2mm] \textit{(Found on Ethereum. Hash: 0xcfb426dcbf8c399d5d9f6f18f8abb60dfe77f42bae51f3cc46425260016c1107. Aug-18-2018 06:56:58 AM +UTC.)} \\[1mm]
    \item ``love makes us fragile, but it is still everyone's greatest wish. The weaker we are, the more we crave it.'' \\[2mm] \textit{(Found on Ethereum. Hash: 0x35afd31eb8cb974405898364c11b86057aa9674a714aded4893419999ff8a649. Dec-31-2018 09:49:00 PM +UTC.)} \\[1mm]
    \item ``Hi, Bro. I admire what you're doing. Salute  0xE0E70fDF0D44DD231C1bc522F2885aD85F43b970  This is my address. I hope you can tip me.Thank Bro'' \\[2mm] \textit{(Found on Ethereum. Hash: 0x1deed99febea575059825d5f98fc005846c0d5688598648ff4b940edcac8fe6f. Aug-10-2021 04:10:27 PM +UTC.)} \\[1mm]
    \item ``0xI want to return \$100000 to you next year but you threaten me so I won't pay you anymore'' \\[2mm] \textit{(Found on Ethereum. Hash: 0x96ec1b4c820a32d648a8251f494985f178532945cc60e484477ba421cfae11ae. Dec-05-2020 04:28:24 PM +UTC.)} \\[1mm]
    \item ``AK47 payout for deposit ETH  - payout 17 of 50. Thank you!'' \\[2mm] \textit{(Found on Ethereum. Hash: 0x593cabe13352f5d3f9eda42264dae7c79d97906b95b9bc05a23f63826d01be12. May-04-2018 11:17:20 AM +UTC.)} \\[1mm]
    \item ``Yang , happy birthday! Welcome to 21 club!'' \\[2mm] \textit{(Found on Ethereum. Hash: 0x6e557b1d3c6ff17b44bc213064e2980f4d7c819f14a3fca1fcd58f472bf8c5af. Jul-10-2021 11:49:07 PM +UTC.)} \\[1mm]
    \item ``If you read this you are a real crypto lover, welcome in the (eternal) KRYPTOSPHERE!'' \\[2mm] \textit{(Found on Ethereum. Hash: 0x9d377734ffcb51efe1d7a828c4c8ca9e7bfe70f4bfcfb6a40380f7cff3175c70. Sep-23-2021 04:27:27 PM +UTC.)}
\end{itemize}

\end{appendix}

\end{document}